\documentclass[runningheads]{paper478-llncs}
\usepackage{latexsym}
\usepackage{makecell}
\usepackage{amsmath}
\usepackage[misc]{ifsym}
\newcommand\blfootnote[1]{%
  \begingroup
  \renewcommand\thefootnote{}\footnote{#1}%
  \addtocounter{footnote}{-1}%
  \endgroup
}
\usepackage{times}
\usepackage{helvet}
\usepackage{courier}
\usepackage{booktabs}
\usepackage{multirow}
\usepackage{multicol}
\usepackage{amssymb}
\usepackage{subfigure}
\usepackage{hyperref}
\usepackage{array}
\usepackage{comment}
\usepackage{graphicx}
\usepackage{comment}
  
\begin{document}
\title{Boundary-aware Transformers for\\ Skin Lesion Segmentation}
%
%
\author{Jiacheng Wang\inst{1},
        Lan Wei\inst{2}, 
        Liansheng Wang\inst{1}\textsuperscript{(\Letter)},
        Qichao Zhou\inst{3}\textsuperscript{(\Letter)}, \\
        Lei Zhu\inst{4}, 
        Jing Qin\inst{5}}
\authorrunning{Wang and Wei et al.}
%
%
\institute{Department of Computer Science at School of Informatics, Xiamen University \\ \email{jiachengw@stu.xmu.edu.cn,lswang@xmu.edu.cn} \and School of Electrical and Computer Engineering, Xiamen University Malaysia \\\email{weilanlan21@gmail.com} \and Manteia Technologies Co.,Ltd \email{zhouqc@manteiatech.com} \and Department of Computer Science and
Engineering, The Chinese University of Hong Kong \email{lzhu@cse.cuhk.edu.hk} \and Center for Smart Health, School of Nursing, The Hong Kong Polytechnic University \email{harry.qin@polyu.edu.hk}}
\maketitle              
\begin{abstract}
Skin lesion segmentation from dermoscopy images is of great importance for improving the quantitative analysis of skin cancer.\blfootnote{\textsuperscript{} J. Wang and L. Wei---Contributed equally; L. Wang and Q. Zhou---Corresponding authors.} 
However, the automatic segmentation of melanoma is a very challenging task owing to the large variation of melanoma and ambiguous boundaries of lesion areas. 
While convolutional neutral networks (CNNs) have achieved remarkable progress in this task, most of existing solutions are still incapable of effectively capturing global dependencies to counteract the inductive bias caused by limited receptive fields.
Recently, transformers have been proposed as a promising tool for global context modeling by employing a powerful global attention mechanism, but one of their main shortcomings when applied to segmentation tasks is that they cannot effectively extract sufficient local details to tackle ambiguous boundaries.
We propose a novel boundary-aware transformer (BAT) to comprehensively address the challenges of automatic skin lesion segmentation.
Specifically, we integrate a new boundary-wise attention gate (BAG) into transformers to enable the whole network to not only effectively model global long-range dependencies via transformers but also, simultaneously, capture more local details by making full use of boundary-wise prior knowledge.
Particularly, the auxiliary supervision of BAG is capable of assisting transformers to learn position embedding as it provides much spatial information.
We conducted extensive experiments to evaluate the proposed BAT and experiments corroborate its effectiveness, consistently outperforming state-of-the-art methods in two famous datasets\blfootnote{\textsuperscript{} Code is available at \url{https://github.com/jcwang123/BA-Transformer}}. 

\end{abstract}
\section{Introduction}
Melanoma is one of the most rapidly increasing cancers all over the world. 
According to the American Cancer Society’s estimation, there are about 100,350 new cases and over 65,00 deaths in 2020~\cite{mathur2020cancer}. 
Segmenting skin lesions from dermoscopy images is a key step in skin cancer diagnosis and treatment planning.
In current clinical practice, dermatologists usually need to manually delineate skin lesions for further analysis.  
However, manual delineation is usually tedious, time-consuming, and error-prone. 
To the end, automated segmentation methods are highly demanded in clinical practice to improve the segmentation efficiency and accuracy.
\begin{figure*}[t]
    \includegraphics[width=\textwidth]{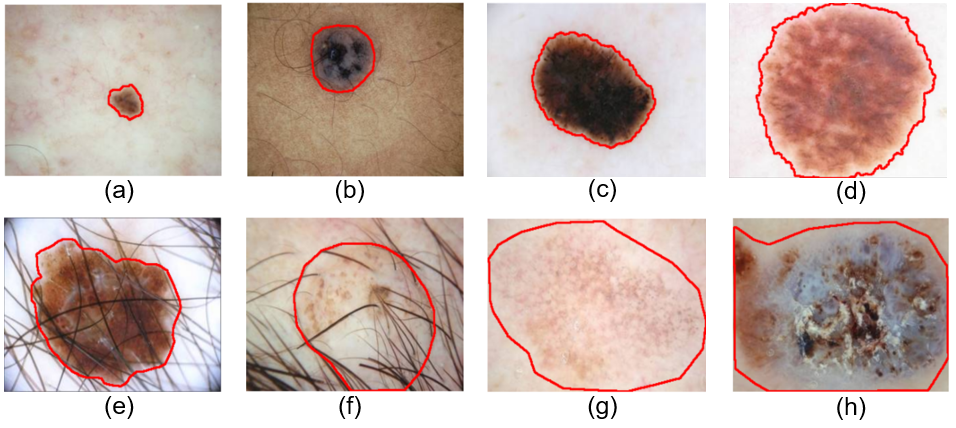}
    \caption{The challenges of automatic skin lesion segmentation from  dermoscopy images: (a)-(d) large skin lesion variations in size, shape, and color, (f)(g) partial occlusion by hair, and (f)-(h) ambiguous boundaries.
    }
    \label{dataset}
\end{figure*}
It remains, however, a very challenging task because (1) skin lesions have large variations in size, shape,and color (see Fig.~\ref{dataset} (a-d)), (2) present of hair will partially cover the lesions destroying local context, (3) the contrast between some lesions to normal skin are relatively low, resulting in ambiguous boundaries (see Fig.~\ref{dataset} (e-h)), and (4) the limited training data make the task even harder.
A lot of effort has been dedicated to overcoming these challenges.
Traditional methods based on various hand-crafted features are usually not stable and robust, leading to poor segmentation performance when facing lesions with large variations~\cite{5342426}. 
The main reason is that these hand-crafted features are incapable of capturing distinctive representations of skin lesions.
To solve the problem, deep learning models based on convolutional neural networks (CNN) have been proposed and achieved remarkable performance gains compared with traditional methods such as some advanced version of the fully convolutional network (FCN)~\cite{yuan2017automatic,7792699}.
However, these models are still insufficient to tackle the challenges of skin lesion segmentation due to the inductive bias caused by the lack of global context. 
With regard to this, researchers propose various approaches to enlarging the receptive fields inspired by the advancement of dilated convolution~\cite{yu2017dilated,yu2019multi}.
Lee~\textit{et al}~\cite{9157193} extensively incorporate the dilated attention module with boundary prior so that the network predict boundary key-points maps to guide the attention module.

Nevertheless, most existing solutions are still incapable of effectively capturing sufficient global context to deal with above mentioned challenges. 
Recently, \emph{transformers} have been proposed to regard an image as a sequence of patches and aggregate feature in global context by self-attention mechanisms~\cite{carion2020detr,prangemeier2020attention}. 
For example, TransUNet~\cite{chen2021transunet}, a hybrid architecture of CNN and transformer, performs well on Synapse multi-organ segmentation.
Yet, it is difficult for transformer based framework to achieve the same success on skin lesion segmentation, which usually has only thousands of data not the same as what they have done in the COCO 2017 Challenge~\cite {carion2020end} containing 118k training images and 5k validation images.
Limited images make it difficult to encode position embedding, and hence will not always be able to accurately and effectively model long-range interactions.
Moreover, regions of lesion cover a relatively small area compared to normal tissues and generally has ambiguous boundary not as human organs, which interferes with segmentation performance by a large margin.

In this paper, we propose boundary-aware transformer (BAT) to ably handle aforementioned problems, by holistically leveraging the advancement of boundary-wise prior knowledge and transformer-based network. 
In fact, this design is based on the intuitions for human beings to perceive lesions in vision, i.e. considering global context to coarsely locate lesion area and paying special attention to ambiguous area to specify the exact boundary.
Concretely, we propose a boundary-wise attention gate (BAG) in transformer architecture to make full use of boundary-wise prior knowledge.
Firstly, BAG would learn which patches in the sequence belong to ambiguous boundary, thus providing a patch-wise attention map to guide this attention gate.
Secondly, a novel key-patch map generation algorithm is introduced for adeptly giving the ground-truth label that can best represent the ambiguous boundary of target lesion.
Thirdly, the auxiliary supervision of BAG provides feedback to train transformers that can let it efficiently learn position embedding on a relatively small dataset.  
We evaluate our model on different publicly available databases. One is the ISBI 2016 and PH2 dataset following the experimental setting in most recent work~\cite{9157193}, the other one is the latest 
ISIC 2018 dataset consisting of 2594 labeled images in total. All the experiment results demonstrate the significant performance gains of our proposed framework.

\section{Method}

\begin{figure*}[t]
    \includegraphics[width=\textwidth]{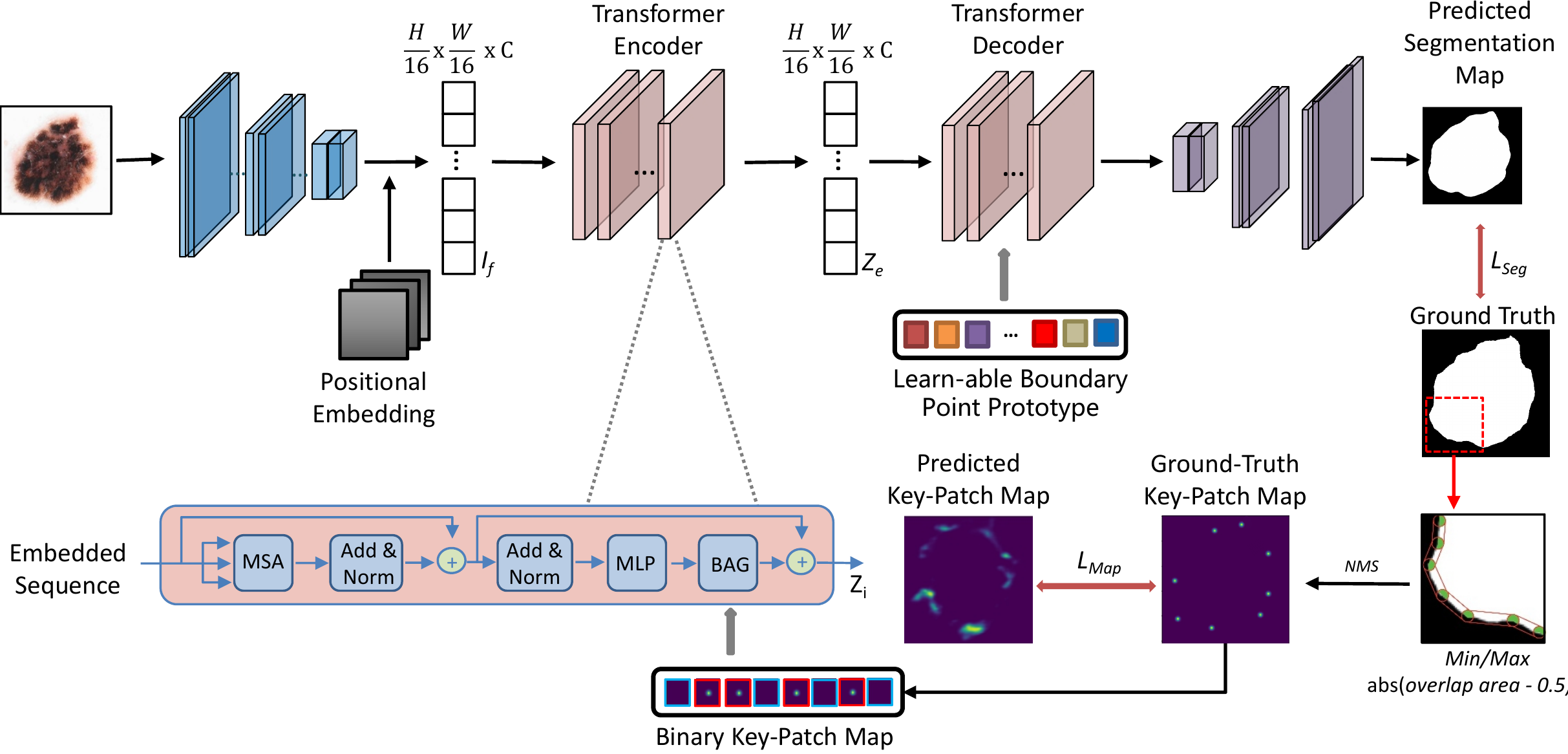}
    \caption{An overview of the proposed Boundary-Aware Transformer framework.}
    \label{framework}
\end{figure*}

An overview of our proposed model is illustrated in Fig.~\ref{framework}.
We will first introduce our basic transformer network that leverages the intrinsic local locality of CNN and innate long-range dependency to complement the skin lesion segmentation in Section~\ref{section:trans_for_seg}.
Then, the overall framework of our boundary-aware transformer for segmenting regions of ambiguous boundary will be elaborated in Section~\ref{section:boundary_prior}. 
In the end, Section~\ref{section_objective_function} introduces the hybrid objective function to efficiently train our boundary-aware transformer.

\subsection{Basic Transformer for Segmentation}
\label{section:trans_for_seg}
Our basic transformer (BT) blends typical CNN's and transformer's architectures to perform segmentation task, by  three procedures: image sequentialization, sequence transformation, and atrous prediction. Image sequentialization is typically the primary step at which an image is converted into 1D sequential embedding fitting the input format of subsequent sequence transformation. 
Atrous prediction is designed to effectively remedy the local feature representation which has been ignored in sequence transformation and robustly segment lesions at multiple scales.

\textbf{Image Sequentialization.}
Current literature processing sequentialization by typical CNN encoder~\cite{carion2020detr,li2020sttr} or linear projection~\cite{dosovitskiy2020vit,zheng2020rethinking}.
Despite its successful advancement achieved on some natural vision tasks, linear projection remains the shortcoming that is the high dependency on the amount of dataset, leading to inferior segmentation performance in medical domain~\cite{chen2021transunet}. 
To this end, we perform sequentialization through the former.
Given an input image $I \in \mathbb{R}^{H \times W\times 3}$ with width $W$ and height $H$, the CNN backbone (ResNet50 as default in this work) produces the corresponding image feature map $I_{f} \in \mathbb{R}^{\tfrac{H}{16} \times \tfrac{W}{16}\times C}$, that the size of each patch is $16\times16$. 
The feature map is then flattened into 1D patch embedding and added by a learnable positial embedding~\cite{gehring2017convolutional} which is randomly initialized to compensate spatial information destroyed by sequentialization, resulting in final sequential embedding as $E \in \mathbb{R}^{L \times C}, L=\tfrac{HW}{256}$.

\textbf{Sequence Transformation.}
Transformer encoder composed of $n$ stacked encoder layers is applied to capture long-range context in a whole dermoscopic image. 
Each layer in the encoder consists of a multi-head self-attention module (MSA) and a Multilayer Perceptron (MLP) following typical design~\cite{vaswani2017attention}.
Assumed that the input of $i$-th layer is $Z^{i-1}$ (specially, $Z^0 \leftarrow E$), the output can be written as follows:
\begin{equation}
    Z^{i} = MSA(Z^{i-1}) \oplus MLP(MSA(Z^{i-1})),
\end{equation}
Eventually, transformed feature of the last layer $Z^{n}$ will be reshaped to 2D format as $Z  \in \mathbb{R}^{\tfrac{H}{16} \times \tfrac{W}{16}\times C}$, for dense prediction in the next.

\textbf{Atrous Prediction.} For segmenting lesions at multiple scales, this module takes transformed feature $Z$ after self-attention mechanism as input and aims to produce a dense prediction. Aiming to enhance the local feature representation and handle the multi-scale lesion context, an atrous prediction module is designed as follows:
\begin{equation}
    \hat{S}_{pred} = \delta(d_{1}^{1}([d_{1}^{3}(Z),d_{3}^{3}(Z),d_{6}^{3}(Z)])).
\end{equation}
Here, $d_r^s(\cdot)$ denotes dilated convolution function with a dilation rate $r$ and filter size of $s \times s$. $\delta$ is a sigmoid function. The enhanced feature maps $d_r^s(Z)$ with various receptive fields are concatenated across channel-wise and projected into segmentation map space. 

\subsection{Boundary-Aware Transformer}
\label{section:boundary_prior}
Efforts to incorporate structural boundary information to CNNs have been made a lot these years, but there is little literature investigating the effectiveness on transformer.
We argue that the equipment of boundary information can also let transformer obtain more power in addressing lesions with ambiguous boundary. 
To this end, we devise the boundary-aware transformer (BAT), in which a boundary-wise attention gate (BAG) is added at end of each transformer encoder layer to refine transformed feature. BAG's architecture is similar to conventional spatial attention gate including (1) a key-patch map generator which takes the transformed feature as input and output a binary patch-wise attention map $\hat{M}_{pred} = \delta(d_1^1(Z))\in \mathbb{R}^{L \times 1}$, where value $1$ indicates that the corresponding patch is at ambiguous boundary. (2) and a residual attention scheme for preserving boundary-wise information. Hence, the boundary-aware transformed feature can be re-written as:
\begin{equation}
\begin{aligned}
    V^{i-1} &= MSA(Z^{i-1}) \oplus MLP(MSA(Z^{i-1})),\\
    Z^{i} &= V^{i-1} \oplus (V^{i-1} \otimes \hat{M}^{i-1}),
\end{aligned}
\end{equation}
where $\oplus$ and $\otimes$ denote element-wise addition and channel-wise multiplication, respectively. 

In addition to BAGs in transformer encoder layers, a query embedding based BAG is applied after encoder to refine the feature $Z^{n}$. It plays the same role in boundary-wise attention but comes true by a totally different way.
Here, instead of learning the linear projection as classifier, we refer a learnable embedding $Q_b$ as context prototype for regions among ambiguous boundary. It will be compared with all patch embedding ($Z$) after aforementioned blocks, to produce a similarity map $M^n$. Those patches with high similarity will be the regions of ambiguous boundary. Similar to other BAGs, a residual attention scheme is also applied here as: $Z^{n+1} = Z^{n} \oplus (Z^{n} \otimes \hat{M}^{n})$. 

By this design, BAT learns robust feature representation of ambiguous boundary in a variety of ways, which is of great significance to handle segmentation of lesions with ambiguous boundary. 
Following our basic design, feature $Z^{n+1}$ is fed into atrous prediction module to produce the segmentation map $\hat{S_{pred}}$.

\textbf{Boundary-Supervised Generator.} 
As the generator doesn't necessarily know on its own which patches can best represent structural boundary of target lesion, we introduce a novel algorithm to produce ground-truth key-patch map to train the generator with full supervision.
Besides the enhancement of boundary features, this design can also help in accelerating training transformer thanks to the auxiliary constraints.

Specifically, boundary points set is produced using conventional edge detection algorithm at first. 
For each point in this set, we draw a circle of radius $r$ (set to $10$ as default) and calculate the proportion $p$ of lesion area in this circle.
Larger or smaller proportion indicates that boundary is not smooth in this cricle.
Thus we score each point as $|p-0.5|$, representing the assistance in segmenting ambiguous parts.
Non-maximum suppression is then utilized to filter points with larger proportion than neighbour $k$ (set to $30$ as default) points.
Next, filtered points' 2D location $(x,y)$ is mapped into 1D location as $\lfloor x/16 \rfloor *16 + \lfloor y/16 \rfloor$, and patch labels at these location are set to $1$ and others are set to 0, leading to final ground-truth $M_{GT}$.

\subsection{Objective Function}
\label{section_objective_function}
To train the segmentation network including the proposed BAGs, we emply two types of loss functions. The first one is a Dice loss function to minimize the difference between the groud-truth segmentation map and the predicted segmentation map as $L_{Seg}$. The second one is a Cross-Entropy loss to reduce the predicted key-patch map and its ground-truth as $L_{Map}$. Total loss is defined as:
\begin{equation}
    \begin{aligned}
        L_{Total} = &L_{Seg} + \sum_{i=1}^{n+1}L_{Map}^{i}, \\
        L_{Seg} = \phi_{DICE}(S_{GT},\hat{S}_{Pred})&,
        L_{Map}^i = \phi_{CE}(M_{GT},\hat{M}_{Pred}^i),
    \end{aligned}
\end{equation}
where $M_{Pred}^i$ denotes the predicted key-patch map at $i$-th transformer encoder layer. $\phi_{DICE},\phi_{CE}$ denote Dice loss function and Cross-Entropy loss function, respectively. $n$ denotes the number of transformer encoder layers and is set to 4 as default. 

\section{Experimental Results}
\subsection{Datasets}
We conduct extensive experiments on the skin lesion segmentation datasets from International Symposium on Biomedical Imaging (ISBI) of the years 2016 and 2018. The datasets are collected from a variety of different treatment centers, archived by the International Skin Imaging Collaboration (ISIC), which hosted a challenge named skin lesion analysis toward melanoma detection to boost the performance of melanoma diagnosis. 
ISIC 2016 contains a total number of 900 samples for training and a total number of 379 dermoscopy images for testing. We follow up the same experimental protocols in the most recent work~\cite{9157193}, in which we train our model on the training set of ISIC 2016 and extensively evaluate it on PH2 dataset.
ISIC 2018 contains 2594 training samples in total and annotation of its public test set is missing, therefore we perform five-fold cross-validation on its training set for fair comparison.

\subsection{Implementation Details}
Our network is implemented on a single NVIDIA RTX 3090Ti. All images are empirically resized to $(512\times512)$ considering the efficiency, and we do data augmentation including vertical flip, horizontal flip and random scale change (limited 0.9-1.1). Each mini-batch includes 24 images and we utilize Adam with an initial learning rate of 0.001 to optimize the network. Learning rate decrease in half when loss on the validation set has not dropped by 10 epochs. The encoder of each network has been pre-trained on ImageNet and all parameters are then fine-tuned for 500 epochs in total.

\begin{table*}[t]
    \caption{Experimental results on different datasets.}\label{table:summary}
\centering
\renewcommand\arraystretch{1.2}
\begin{tabular}{p{2cm}<{\centering}p{1.5cm}<{\centering}p{1.5cm}<{\centering}p{2cm}<{\centering}p{1.5cm}<{\centering}p{1.5cm}<{\centering}}
\hline
\hline
\multirow{2}{*}{Model} & \multicolumn{2}{c}{ISIC 2016 + PH2} & \multirow{2}{*}{Model} & \multicolumn{2}{c}{ISIC 2018}\\
 & $Dice\uparrow$ & $IoU\uparrow$ &  & $Dice\uparrow$ & $IoU\uparrow$ \\
\hline
SSLS\cite{ahn2015automated}  & $0.783$& $0.681$ & DeepLabv3~\cite{chen2017rethinking} & $0.884$& $0.806$\\
MSCA\cite{bi2016automated}  & $0.815$& $0.723$ & U-Net++~\cite{zhou2018unetpp} & $0.879$& $0.805$\\
FCN~\cite{long2015fully}   & $0.894$& $0.821$ & CE-Net~\cite{gu2019cenet} & $0.891$& $0.816$\\
Bi~\textit{et al}~\cite{7942129}  & $0.906$& $0.839$ & MedT~\cite{valanarasu2021medical} & $0.859$ & $0.778$\\
Lee~\textit{et al}~\cite{9157193}   & $0.918$& $0.843$ &TransUNet~\cite{chen2021transunet} & $0.894$& $0.822$\\
\hline
BAT & $0.921$& $0.858$ & BAT & $0.912$& $0.843$\\
\hline
\hline
\end{tabular}
\label{tab:results}
\end{table*}

\begin{figure}[t]
\centering
\includegraphics[width=0.8\linewidth]{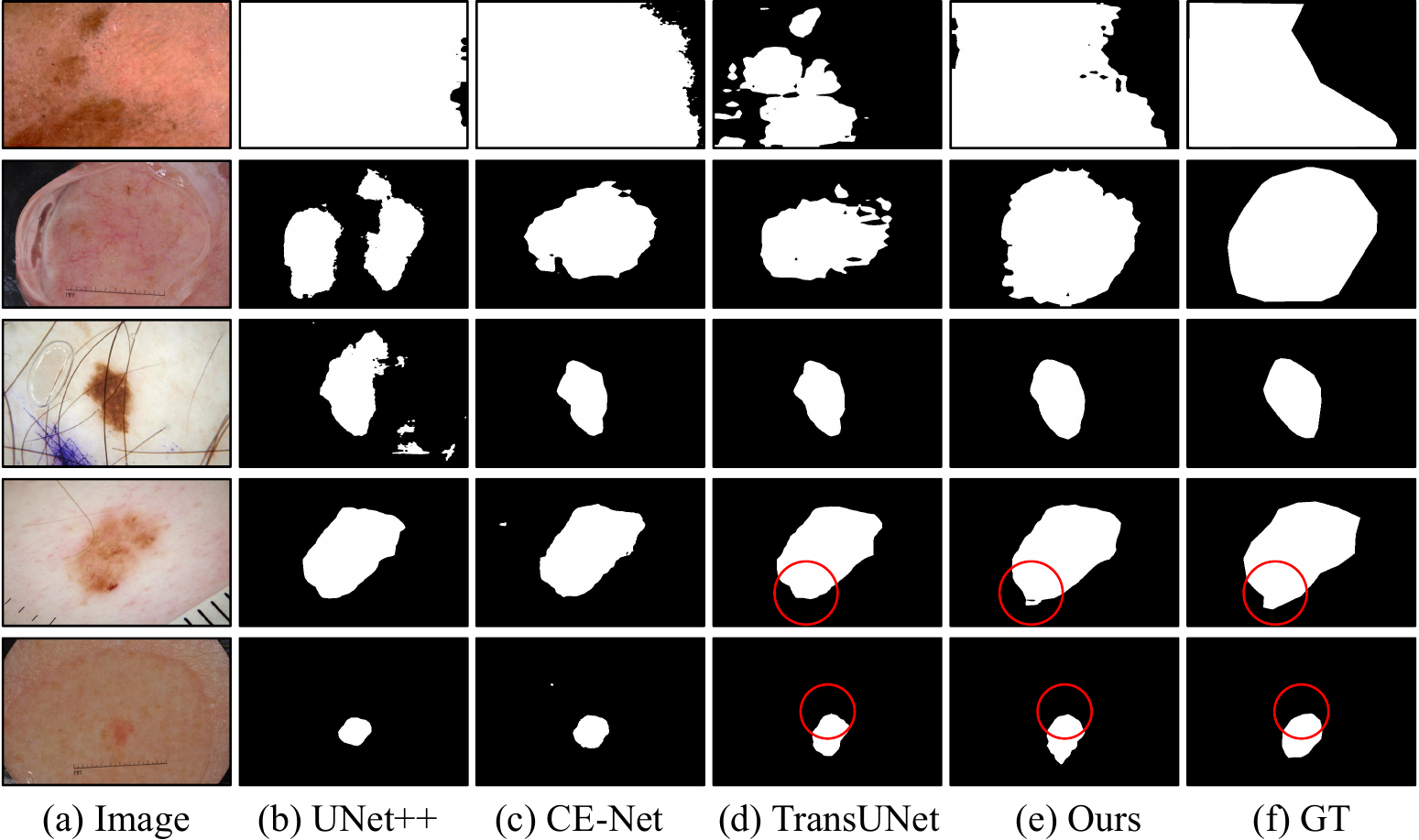}
\caption{Visual comparison of lesion segmentation results produced by different methods.}
\label{fig:comparison_result}
\end{figure}

\subsection{Comparison with State-of-the-Arts}
For baseline comparisons, we run experiments on both convolutional and transformer-based methods.
With regard to the evaluation metrics, we employ a Dice coefficient (Dice), and a Intersection over Union (IoU). Table~\ref{table:summary} displayed the comparative study of our proposed boundary-aware transformer (BAT) with other methods on different datasets. It's obviously shown that our model achieves the best segmentation performance.

On the \emph{ISIC 2016 + PH2} dataset, we compare our method with five state-of-the-art methods. Among them, Lee~\textit{et al}~\cite{9157193} is a 2D attention-based model with use of boundary-prior knowledge, achieving best segmentation performance on skin lesion segmentation recently. As seen in the Table~\ref{table:summary}, our BAT achieves 0.920 in Dice and 0.858 in IoU, outperform Lee~\textit{et al} by 0.2\% and 1.5\% in Dice and IoU, respectively.
We extensively conduct experiments with other SOTA segmentation networks on the \emph{ISIC 2018} dataset, including three famous convolutional models for segmentation (DeepLabv3~\cite{chen2017rethinking}, UNet++~\cite{zhou2018unetpp}, CE-Net~\cite{gu2019cenet}) and two transformer-based network to address medical image segmentation (TransUNet~\cite{chen2021transunet}, MedT~\cite{valanarasu2021medical}). Even compared with other state-of-the-art segmentation models, our BAT still achieves the consistent and significant improvement on both metrics.
It's noteworthy that transformer-based network has superior performance than conventional CNNs, indicating the effectiveness of utilizing global context to detect skin lesion. 
In addition, compared with TransUNet, our method leveraging the boundary-prior knowledge significantly improves the segmentation performance (1.8\% on Dice and 2.1\% on IoU), proving that the combination of boundary information and transformer architecture is indeed helpful to segment target lesion.

Fig.~\ref{fig:comparison_result} visualizes five typical challenging cases of lesion segmentation results. It is observed that our results are closest to the ground truth, when compared with our competitors. The first three rows represent cases with various color, size and shape, and our BAT outperforms others with most stable segmentation performance, indicating the robust advancement of global context. The last two rows highlight some small regions of ambiguous boundary and it's shown that our BAT is capable of tacking such problems, due to the use of boundary-wise prior knowledge.

\subsection{Ablation Study}
\begin{table*}[t]
    \caption{Experimental results on different datasets.}\label{table:ablation}
\centering
\renewcommand\arraystretch{1.2}
\begin{tabular}{p{1.8cm}<{\centering}p{1.8cm}<{\centering}p{1.5cm}<{\centering}p{1.5cm}<{\centering}p{1.5cm}<{\centering}p{1.5cm}<{\centering}}
\hline
\hline
\multirow{2}{*}{Trans.} & \multirow{2}{*}{BAG} & \multicolumn{2}{c}{ISIC 2016 + PH2}& \multicolumn{2}{c}{ISIC 2018}\\
 & & $Dice\uparrow$ & $IoU\uparrow$ &  $Dice\uparrow$ & $IoU\uparrow$ \\
\hline
 & &   $0.884$ & $0.805$ &  $0.879$ & $0.810$\\
\checkmark &   & $0.900$ & $0.827$ &  $0.890$ & $0.821$\\
\checkmark & \checkmark & $0.921$& $0.858$ &  $0.912$& $0.843$\\
\hline
\hline
\end{tabular}
\label{tab:ablation}
\end{table*}
We further conduct ablation studies to demonstrate the effectiveness of three major components in BAT: (1) the transformer-based self-attention mechanism (Trans.), (2) boundary-wise attention gate (BAG). As shown in the Table.~\ref{table:ablation}, by the incorporation of self-attention mechanism, the IoU increases by a large margin on both datasets. This result indicate that it's essential to integrate global context to improve the skin lesion detection.
On the other hand, applying BAGs to guide transformer further improves the performance significantly, confirming the effectiveness of boundary-wise prior knowledge to tackling challenging cases, such as lesions with ambiguous boundary. 

\section{Conclusion}
We present a novel and efficient context-aware network, namely boundary-aware transformer (BAT) network, for accurate segmentation of skin lesion from  dermoscopy images. Extensive experiments on two public datasets confirm the effectiveness of our proposed BAT, to help yield much better segmentation results for skin lesions. Our full model outperforms state-of-the-art models by a large margin in segmentation accuracy and the intuitive visualization shows that our BAT has most satisfactory performance on skin lesions with ambiguous boundary.
\bibliographystyle{paper478}
\bibliography{paper478}
\end{document}